\documentclass[twocolumn,twoside,slac_two]{revtex4}
\usepackage{graphicx}
\usepackage{fancyhdr}
\usepackage{xspace}

\newcommand{\ksn}{\ensuremath{\mathrm{K_S}}\xspace}
\newcommand{\ksln}{\ensuremath{\mathrm{K_{S,L}}}\xspace}
\newcommand{\kln}{\ensuremath{\mathrm{K_L}}\xspace}

\newcommand{\kn}{\ensuremath{\mathrm{K^0}}\xspace}
\newcommand{\knb}{\ensuremath{\mathrm{\bar{K}^0}}\xspace}
\newcommand{\bn}{\ensuremath{\mathrm{B^0}}\xspace}
\newcommand{\bnb}{\ensuremath{\mathrm{\bar{B}^0}}\xspace}

\begin{document}

\title{CPT and QM tests using kaon interferometry}

%

\author{A. Di Domenico}
\affiliation{Dipartimento di Fisica, Universit\`a di Roma 
``La Sapienza'', \\
and I.N.F.N. Sezione di Roma, 
P.le A. Moro, 2, I-00185 Rome, Italy \\
e-mail: antonio.didomenico@roma1.infn.it}
%

\begin{abstract}
The neutral kaon system offers a unique possibility to perform fundamental tests of $CPT$ invariance, as well as of the basic
principles of quantum mechanics. The most recent and significant 
limits on $CPT$ 
violation are reviewed, including the ones related to possible 
decoherence mechanisms
or Lorentz symmetry breaking, which
might be induced by quantum gravity.
The experimental results show no
deviations from the expectations of quantum mechanics and $CPT$ symmetry, while the accuracy in some cases reaches the interesting 
Planck scale region. 
Finally, prospects for this kind of experimental studies
at the upgraded DA$\Phi$NE 
$e^+e^-$ collider at Frascati are briefly
discussed.
\end{abstract}

\maketitle

\thispagestyle{fancy}


\section{Introduction}
The three discrete symmetries of quantum mechanics, $C$ (charge conjugation), 
$P$ (parity), and $T$ (time reversal) are known to be violated 
in nature, both singly and in pairs.
Only the combination of the three --- $CPT$ (in any order) --- 
appears to be an exact symmetry of nature.
\par
A rigorous proof of the $CPT$ theorem can be found in Refs
\cite{luders,pauli,bell,jost} (see also Refs \cite{greenberg1,greenberg2,hollands} for some recent developments);
it ensures that exact $CPT$ 
invariance holds for any quantum field theory
assuming (1) Lorentz invariance, (2) Locality, and (3) Unitarity 
(i.e. conservation of probability).
Testing the validity of $CPT$ invariance therefore probes 
the most fundamental
assumptions of our present understanding of particles 
and their interactions.

\par
The neutral kaon doublet is one of the most intriguing systems in nature.
During its time evolution a neutral kaon oscillates between its particle and 
antiparticle states with a beat frequency 
$\Delta m 
\approx 5.3 \times 10^9 
\hbox{ s}^{-1}$, where $\Delta m$ is the small mass difference between the
exponentially decaying states \kln and \ksn. 
%
%
%
The fortunate coincidence that 
$\Delta m$  
is about half the 
decay width of \ksn makes it possible to observe a variety 
of intricate interference phenomena in the production and decay 
of neutral kaons. In turn, such observations enable us to test 
quantum mechanics, the interplay of different conservation laws and the validity of various symmetry principles. 
In particular 
the extreme sensitivity of the neutral kaon system to 
a variety of $CPT$-violating effects makes it one of the best candidates
for an accurate experimental test of 
this
symmetry \cite{didohand}.
As a figure of merit, 
the fractional mass 
difference $\left({m_{\kn}-m_{\knb}}\right)/{m_{\kn}}$ can be considered:
it can be measured at the level of $\mathcal{O}(10^{-18})$ for neutral kaons,
while, for comparison, a limit 
of $\mathcal{O}(10^{-14})$ can be reached on the corresponding quantity 
for the $\bn-\bnb$ system,
 and only of $\mathcal{O}(10^{-8})$ for proton-antiproton \cite{pdg}.
\section{CPT test from unitarity}
\par The real part of the complex parameter $\delta$, describing $CPT$ violation
 in $\kn-\knb$ mixing, has been
 measured by the CPLEAR collaboration studying the time behaviour of semileptonic decays
from initially tagged \kn and \knb mesons \cite{cplearred}:
%
%
\begin{eqnarray}
  \label{eq:red_cplear}
\Re\delta=(0.30 \pm 0.33_{\mbox{stat}} \pm 0.06_{\mbox{syst}})\times 10^{-3}~.
\end{eqnarray}
\par
One of the most precise and significant tests of the $CPT$ symmetry comes from the unitarity relation, originally derived by Bell and Steinberger \cite{BS}: 
\begin{eqnarray}
&& \left( \frac{\Gamma_S+\Gamma_L}{\Gamma_S-\Gamma_L}+i \tan \phi_{SW}\right) 
\left[
\frac{\Re \epsilon}{1+|\epsilon|^2} - i \Im \delta
\right]  \nonumber \\
& &= \frac{1}{\Gamma_S-\Gamma_L} \sum_f A^*(K_S\rightarrow f)A(K_L\rightarrow f)  \nonumber \\
& & \equiv\sum_f \alpha_f~,
\label{eq:bs}
\end{eqnarray}
where $\epsilon$ is the usual complex parameter describing
$CP$ violation in $\kn-\knb$ mixing; $\Gamma_S$ and $\Gamma_L$
are the widths of the physical states \ksn and \kln; $\phi_{SW}$ is the {superweak} phase; $A(K_i\rightarrow f)$ is the decay amplitude of the state $K_i$
into final state $f$, 
and the sum runs over all possible final states.
The above relationship can be used to bound the parameter
$\Im{\delta}$,
after having provided all the $\alpha_i$ parameters, $\Gamma_S$, $\Gamma_L$, and $\phi_{SW}$ as inputs. 
Using several measurements from the KLOE experiment \cite{kloebs}, 
 values from the Particle Data Group (PDG), and a combined fit of KLOE and CPLEAR data, the following result is obtained \cite{pdg}:
\begin{eqnarray}
\Im \delta = (-0.6 \pm 1.9)\times 10^{-5} ~,
\end{eqnarray}
which is the most stringent limit on $\Im\delta$\footnote{The result $\Re\epsilon=(161.2 \pm 0.6)\times 10^{-5} $, which is obtained in the same analysis, 
is not relevant for the discussion here.}, 
the main limiting factor of this result being the uncertainty on the phase
$\phi_{+-}$ 
entering in the parameter $\alpha_{\pi^+\pi^-}$.
\par
    The limits on $\Im\delta$ and $\Re\delta$ 
can be used
   to constrain the mass and width difference between \kn
   and \knb.
In the limit $\Gamma_{\kn}-\Gamma_{\knb}=0$, i.e. 
neglecting $CPT$-violating effects in the decay amplitudes,
the best bound on the neutral kaon mass difference is obtained:
$$\left|{m_{\kn}-m_{\knb}}\right|< 5.1  \times 10^{-19}~ {\rm GeV 
\quad at~ 95 ~\% ~CL}~. $$
A preliminary update including the latest results on $\phi_{+-}$ 
by the KTeV collaboration \cite{ktevepsp} yields slightly improved results 
\cite{palutan08}:
\begin{eqnarray}
&\Im \delta = (-0.1 \pm 1.4)\times 10^{-5} &\nonumber \\
&\left|{m_{\kn}-m_{\knb}}\right|< 4.0  \times 10^{-19}~ {\rm GeV 
\quad at~ 95 ~\% ~CL}~.& \nonumber
\end{eqnarray}

\section{CPT and QM tests}
\par DA$\Phi$NE, the Frascati $\phi$-factory, is an $e^+ e^-$ collider 
working at a center of mass energy of $\sqrt s \sim 1020$ MeV, 
corresponding to the peak of the $\phi$ resonance. The $\phi$
production cross section is $\sim 3 \mu \mbox{b}$, and its 
decay into $K^0 \bar{K^0}$ 
has a branching fraction of $34 \%$.
The neutral kaon pair is produced in a coherent quantum state with quantum numbers $J^{PC}=1^{- -}$:
\begin{eqnarray}
  |i \rangle   =  {1\over \sqrt{2}} \{ |K^0 \rangle |\bar{K}^0 \rangle - 
 |\bar{K}^0 \rangle |K^0 \rangle
\} 
\nonumber\\
= {N\over \sqrt{2}} \{ |K_S \rangle |K_L \rangle - 
 |K_L \rangle |K_S \rangle
\label{eq:state}
\}
\end{eqnarray}  
where $N={\sqrt{(1+|\epsilon_S|^2)(1+|\epsilon_L|^2)}}/{(1-\epsilon_S\epsilon_L)} \simeq 1$
is a normalization factor, and $\epsilon_{S,L}=\epsilon \pm \delta$.
\par
The detection of a kaon at large (small) times {\it tags} a
$K_S$ ($K_L$) in the opposite direction. 
\par
The KLOE detector
consists mainly of a large 
volume drift chamber\cite{dc} surrounded by an electromagnetic calorimeter\cite{emc}. A superconducting coil provides a 0.52 T solenoidal magnetic field. 
\\ At KLOE a $K_S$ is tagged by identifying the interaction 
of the $K_L$ in the calorimeter ($K_L$-crash), while
a $K_L$ is tagged by detecting a $K_S\rightarrow \pi^+\pi^-$ decay near the 
interaction point (IP).
\par
KLOE completed the data taking in March 2006 with a total integrated luminosity 
${\rm L} \sim 2.5 \mbox{~fb}^{-1}$, corresponding to  
$\sim 7.5 \times 10^{9}$ $\phi$-mesons produced.
\par
The quantum interference between the two kaons initially in the 
{\it entangled} state in eq.(\ref{eq:state}) and decaying in
the $CP$ violating channel $\phi\rightarrow \ksn\kln \rightarrow \pi^+\pi^-
\pi^+\pi^-$,
has been observed for the first time by the KLOE collaboration \cite{kloeqm}.
The measured $\Delta t$
distribution, with $\Delta t$ the absolute value of the time
difference of the two $\pi^+\pi^-$ decays,  
can be fitted with the distribution:
\begin{eqnarray}
\label{eq:deckloe}
I(\pi^+\pi^-,\pi^+\pi^-;\Delta t)\propto 
e^{-\Gamma_L \Delta t}+
e^{-\Gamma_S \Delta t}
\nonumber \\
-2 (1-\zeta_{SL})e^{-{{(\Gamma_S+\Gamma_L)}\over{2}}\Delta t}\cos( \Delta m \Delta t)~,
\end{eqnarray}
where the quantum mechanical expression in the $\{\ksn,\kln\}$ basis has been modified with the introduction of a decoherence parameter $\zeta_{SL}$, and a factor $(1-\zeta_{SL})$ multiplying the interference term. Analogously,  a $\zeta_{0\bar{0}}$ parameter can be defined in the $\{\kn,\knb\}$ basis \cite{bertlmann1}.
After having included resolution and detection efficiency effects,
having taken into account the background due to 
coherent and incoherent \ksn-regeneration on the beam pipe wall, the small contamination of non-resonant $e^+e^-\rightarrow\pi^+\pi^-\pi^+\pi^-$ events,
and keeping 
$\Delta m$, $\Gamma_S$ and $\Gamma_L$ fixed to the PDG values, the fit is performed on the $\Delta t$ distribution. The analysis of a data sample corresponding to $\rm L\sim 380\,\mathrm{pb}^{-1}$ yields the following results \cite{kloeqm}:
\begin{eqnarray}
\label{eq:deckloe2}
\zeta_{SL}&=&0.018\pm0.040_{\mbox{\rm stat}}\pm0.007_{\mbox{\rm syst}} \nonumber \\
\zeta_{0\bar{0}}&=&(1.0\pm2.1_{\mbox{\rm stat}}\pm0.4_{\mbox{\rm syst}})\times 10^{-6}~,
\end{eqnarray}
compatible with the prediction of quantum mechanics, i.e. $\zeta_{SL}=\zeta_{0\bar{0}}=0$, and no decoherence effect. 
In particular the result on $\zeta_{0\bar{0}}$ has a high precision,
 $\mathcal{O}(10^{-6})$, due to the $CP$ suppression present in the specific 
decay channel;
it is an improvement by five orders of magnitude over the previous limit,
obtained by 
Bertlmann and co-workers \cite{bertlmann1} in a re-analysis of
CPLEAR data \cite{cplearqm}. This result can also be compared to a similar one recently
obtained in the B meson system \cite{aurelio}, where an accuracy of
$\mathcal{O}(10^{-2})$ has been reached.
\par
At a microscopic level, in a quantum gravity picture, 
space-time might be
subjected to inherent non-trivial quantum metric and topology 
fluctuations
at the Planck scale ($\sim 10^{-33}\hbox{~cm}$), 
called generically {\it space-time foam}, with associated microscopic event horizons.
This space-time structure
would lead to pure states evolving to mixed states, i.e.\ the decoherence
of apparently isolated matter systems \cite{hawk2}.
This decoherence, in turn, necessarily
implies, by means of a 
theorem \cite{wald}, 
$CPT$ violation,
in the sense that the quantum
mechanical operator generating $CPT$ transformations cannot be 
consistently defined.
%
\par 
A model for decoherence can be formulated \cite{ellis1}
in which 
a single kaon
is 
described by a 
density matrix $\rho$ that
obeys  a modified Liouville-von Neumann equation:
\begin{equation}
  \label{eq:evolmod2}
  \frac{d \rho}{dt} = -i{\bf H}\rho  +i\rho {\bf H}^{\dagger} + 
L(\rho; \alpha,\beta,\gamma)
\end{equation}
where 
${\bf H}$ is the
neutral kaon effective Hamiltonian, and
the extra term $L(\rho; \alpha,\beta,\gamma)$
would induce decoherence in the system, and depends on
%
three real parameters, $\alpha, \beta$ and $\gamma$,
which violate $CPT$ symmetry and quantum mechanics (they
satisfy the inequalities
$\alpha$, $\gamma >0$ and $\alpha \gamma > \beta^2$ - see Refs. \cite{ellis1,ellis2}).
%
They have units of mass
and are presumed to be at most
$\mathcal {O} (m^2_K/M_{Planck}) \sim 2 \times 10^{-20} \,\mbox{GeV}$,
where $M_{Planck}=1\sqrt{G_N}= 1.22\times 10^{19} \mbox{~GeV}$ is the Planck mass.
\par
The CPLEAR collaboration, studying the time behaviour of 
single neutral kaon decays to 
$\pi^+\pi^-$ and $\pi e \nu$ final states, obtained the following results
\cite{cplearabc}:
\begin{eqnarray}
  \label{eq:abg_cplear}
  \alpha &=&  \left(-0.5 \pm 2.8 \right) \times 10^{-17} \,\mbox{GeV} \nonumber\\
  \beta &=& \left(2.5 \pm 2.3  \right)\times 10^{-19} \,\mbox{GeV} \nonumber\\
  \gamma &=& \left(1.1 \pm 2.5  \right)\times 10^{-21}\, \mbox{GeV}~.
\end{eqnarray}
\par
The KLOE collaboration, studying the same
$I(\pi^+\pi^-,\pi^+\pi^-;\Delta t)$ 
distribution as in the $\zeta$ parameters analysis,
in the simplifying hypothesis of complete positivity\footnote{This hypothesis,
reducing the number of free parameters, makes the fit of the experimental
distribution easier, even though it is not strictly necessary.}
 \cite{benattiall}, i.e. $\alpha=\gamma$ and
$\beta=0$,
obtained the following
result \cite{kloeqm}:
\begin{eqnarray}
  \label{eq:kloegamma}
  \gamma &=& \left({1.3 ^{+2.8} _{-2.4}}_{\mbox{stat}}  \pm 0.4_{\mbox{syst}} \right)\times 10^{-21}\, \mbox{GeV}~,
\end{eqnarray}
All results are compatible with no $CPT$ violation, while
the sensitivity approaches
the interesting level of 
$\mathcal {O} (10^{-20} \,\mbox{GeV})$. 
\par
As discussed above, 
in a quantum gravity 
framework inducing decoherence, the $CPT$ operator is {\it ill-defined}. 
This consideration might have
intriguing consequences in correlated neutral kaon states, where
the resulting loss of particle-antiparticle identity could induce a 
breakdown of the correlation in state (\ref{eq:state}) 
imposed by Bose statistics \cite{mavro1,mavro2}.
As a result the initial state (\ref{eq:state}) can be parametrized in general as:
\begin{eqnarray}
|i \rangle  & = &  {1\over \sqrt{2}} [ |K^0\rangle |\bar{K}^0\rangle -|\bar{K}^0\rangle |K^0 \rangle  
\nonumber \\
& & + \omega \left( |K^0\rangle |\bar{K}^0\rangle + 
|\bar{K}^0\rangle |K^0\rangle \right) ]~,
\label{eq:state5}
\end{eqnarray}  
where $\omega$ is a complex parameter describing a completely novel $CPT$
violation phenomenon, not included in previous analyses. Its order of magnitude
 could be at most $$|\omega| \sim \left [(m^2_K/M_{\rm{Planck}})/\Delta \Gamma \right ]^{1/2} \sim 10^{-3}$$ with $\Delta \Gamma = \Gamma_S - 
\Gamma_L$. 
A similar analysis performed by the KLOE collaboration 
on the same  $I(\pi^+\pi^-,\pi^+\pi^-;\Delta t)$ 
distribution as before,
including in the fit 
the modified initial state eq.(\ref{eq:state5}), 
yields the first measurement
of the complex parameter $\omega$ \cite{kloeqm}:
\begin{eqnarray}
\label{eq:omega}
\Re(\omega) &=&\left( {1.1^{+8.7}_{-5.3}}_{\mbox{stat}} \pm 0.9_{\mbox{syst}} \right)\times{10^{-4}}
\nonumber \\
\Im(\omega) &=&\left( {3.4^{+4.8}_{-5.0}}_{\mbox{stat}}\pm 0.6_{\mbox{syst}} \right)\times{10^{-4}} ~,
\end{eqnarray}
with an accuracy that already reaches the interesting Planck scale region.
\par
A preliminary analysis of a KLOE data sample corresponding to 
$\rm L \sim 1 \quad fb^{-1}$ yields the following updated results \cite{moriond08}:
\begin{eqnarray}
\label{eq:deckloe2}
\zeta_{SL}&=&0.009\pm0.022_{\mbox{\rm stat}} \nonumber \\
\zeta_{0\bar{0}}&=&\left(0.03\pm1.2_{\mbox{\rm stat}}\right) \times 10^{-6}  \nonumber \\
  \gamma &=& \left( {0.8 ^{+1.5}_{-1.3}}_{\mbox{stat}} \right)\times{10^{-21}}\, \mbox{GeV}  \nonumber \\
\Re(\omega) &=&\left( {-2.5^{+3.1}_{-2.3}}_{\mbox{stat}} \right)\times{10^{-4}}
\nonumber \\
\Im(\omega) &=&\left( {-2.2^{+3.4}_{-3.1}}_{\mbox{stat}} \right)\times{10^{-4}} ~, \nonumber
\end{eqnarray}
while the analysis of the full KLOE data sample is being completed.

\section{CPT violation and Lorentz symmetry breaking}
$CPT$ invariance holds 
for any realistic Lorentz-invariant quantum field theory. 
However a very general theoretical possibility for $CPT$ violation is
based on spontaneous breaking of Lorentz symmetry \cite{kost1,kost2,kost3}, 
which 
appears to 
be compatible with the basic
tenets of quantum field theory and retains the property of gauge
invariance and renormalizability (Standard Model 
Extensions - SME).
In SME for neutral kaons, $CPT$ violation manifests to lowest order only in the parameter $\delta$,
 and exhibits
a dependence on the 4-momentum of the kaon:
%
\begin{eqnarray}
\label{eq:deltak}
\delta \approx i \sin \phi_{SW} e^{i \phi_{SW}} \gamma_K (\Delta a_0-
\vec{\beta_K}\cdot \Delta{\vec{a}})/\Delta m
\end{eqnarray}  
where $\gamma_K$ and $\vec{\beta_K}$ are the kaon boost factor and velocity in the observer frame, and $\Delta a_{\mu}$
are four $CPT$- and Lorentz-violating coefficients for the two valence quarks in the kaon.
\par 
Following Ref.\ \cite{kost2}, the time dependence arising from the rotation of the Earth can be explicitly displayed in eq. (\ref{eq:deltak}) by choosing a three-dimensional basis ($\hat{X},\hat{Y},\hat{Z}$) in a non-rotating frame, with the $\hat{Z}$ axis along the Earth's rotation axis, and a basis $(\hat{x},\hat{y},\hat{z})$ for the rotating (laboratory) frame.
The $CPT$ violating parameter $\delta$ may then be expressed as:
\begin{eqnarray}
\label{eq:deltak3}
\delta &=& \frac{1}{2\pi}\int_0^{2\pi}\delta(\vec{p},t_{sid})d\phi\nonumber\\
&=& {\frac{i \sin \phi_{SW} e^{i \phi_{SW}}}{\Delta m}} 
\gamma_K 
\{
\Delta a_0 \nonumber\\
&&+\beta_K\Delta a_Z \cos\theta\cos\chi \nonumber\\
&&+\beta_K ( \Delta a_Y \sin\chi\cos\theta \sin\Omega t_{sid} \nonumber \\
&& +\Delta a_X \sin\chi\cos\theta \cos\Omega t_{sid}) \} ~,
\end{eqnarray}  
where $t_{sid}$ is the sidereal time, $\Omega$ is the Earth's sidereal frequency, 
$\cos\chi=\hat{z}\cdot\hat{Z}$, $\theta$ and $\phi$
are the conventional polar and azimuthal angles defined in the laboratory 
frame about the $\hat{z}$ axis, and an integration on the 
azimuthal angle $\phi$ has been performed, assuming a symmetric 
decay distribution in this variable\footnote{Although 
not necessary, this assumption is taken here in order to 
simplify formulas.}.
The sensitivity to the four  $\Delta a_{\mu}$
parameters can be very different for fixed target and collider experiments,
showing complementary features \cite{kost2}.
\par 
At KLOE
the $\Delta a_0$ parameter can be evaluated through 
the difference 
of
the semileptonic charge asymmetries:
\begin{eqnarray}
A_{S,L}&=&\frac{\Gamma(\ksln\rightarrow \pi^- l^+ \nu)
-\Gamma(\ksln\rightarrow \pi^+ l^- \bar{\nu})}{\Gamma(\ksln\rightarrow \pi^- l^+ \nu)
+\Gamma(\ksln\rightarrow \pi^+ l^- \bar{\nu})} ~, \nonumber 
\end{eqnarray}
by performing the measurement of each asymmetry with a 
symmetric integration over the polar angle
$\theta$, 
thus averaging to zero 
any possible contribution from the terms
proportional to $\cos\theta$ in eq.(\ref{eq:deltak3}):
\begin{eqnarray}
\label{eq:delta0}
A_S-A_L \simeq \left[{
\frac{4\Re\left( i \sin \phi_{SW} e^{i \phi_{SW}} \right)\gamma_K}{\Delta m}}  \right]
 \Delta a_0 ~.
\end{eqnarray}
In this way a first preliminary evaluation of the $\Delta a_0$ parameter 
can be obtained by KLOE \cite{didohand,didomecpt07}:
\begin{equation}
\Delta a_0 = (0.4\pm 1.8)\times 10^{-17}\hbox{ GeV}~.
\end{equation}
With the analysis of the full KLOE data sample ($L=2.5\hbox{ fb}^{-1}$) an 
accuracy $\sigma(\Delta a_0) \sim 7 \times 10^{-18}\hbox{ GeV}$ could 
be reached.
\par
At KLOE the $\Delta a_{X,Y,Z}$ parameters 
can be evaluated performing a sidereal time dependent analysis
of the asymmetry:
\begin{eqnarray}
\label{eq:deltaz2}
A(\Delta t)= \frac{
N^+-N^-}
{N^++N^-} ~,
\nonumber
\end{eqnarray}
with
\begin{eqnarray}
N^+=I\left(\pi^+\pi^-(+),\pi^+\pi^-(-);\Delta t>0\right) \nonumber \\
N^-=I\left(\pi^+\pi^-(+),\pi^+\pi^-(-);\Delta t<0\right) ~,
\nonumber
\end{eqnarray}
where the two identical final states are distinguished 
 by their emission in the forward ($\cos\theta>0$) or backward 
($\cos\theta<0$)
hemispheres
(denoted by the symbols $+$ and $-$, 
respectively), and $\Delta t$ is the time difference between $(+)$ and $(-)$ $\pi^+\pi^-$ decays.
A preliminary analysis based on a data sample corresponding to an integrated
luminosity $\rm L\sim 1\,\mathrm{fb}^{-1}$ 
yields the following results 
\cite{didohand,didomecpt07,moriond08}:
\begin{eqnarray}
\Delta a_X = (-6.3\pm 6.0)\times 10^{-18}\hbox{ GeV} \nonumber \\
\Delta a_Y = (2.8\pm 5.9)\times 10^{-18}\hbox{ GeV} \nonumber \\
\Delta a_Z = (2.4\pm 9.7)\times 10^{-18}\hbox{ GeV}~.
\end{eqnarray}
\par
A preliminary measurement performed by the KTeV collaboration \cite{ktev3} based on the search
for sidereal time variation of the phase
$\phi_{+-}$ 
constrains $\Delta a_{X}$ and $\Delta a_{Y}$ 
to less than $9.2 \times 10^{-22}\hbox{ GeV}$ at $90\%$ C.L.
These results can also be compared to similar ones recently
obtained in the B meson system \cite{babarlv}, where an accuracy 
on the $\Delta a_{\mu}^B$ parameters of 
$\mathcal{O}(10^{-13}{\rm GeV})$ has been reached.
\section{Future plans}
A proposal \cite{kloe2prop,rollin} has been presented
for a
physics program 
to be carried out with an upgraded KLOE detector, 
KLOE-2, at an upgraded DA$\Phi$NE machine,
which is expected to deliver an 
integrated luminosity up to $20 \div 50$ fb$^{-1}$.
The major upgrade of the KLOE detector would consist in the addition of an inner tracker for the improvement of decay vertex resolution, therefore 
improving the sensitivity on several parameters 
based on kaon interferometry measurements. 
The KLOE-2 program concerning neutral kaon interferometry 
is 
summarized in table \ref{tab:modes1}, where the 
KLOE-2 statistical sensitivities on the main parameters that can be 
extracted from kaon decay
time distributions  
$I(f_1,f_2;\Delta t)$
(with different choices of final states $f_1$ and $f_2$)
are listed 
for an assumed
integrated luminosity $L = 50$ fb$^{-1}$, and compared 
to the best presently published measurements.

\begin{table*}[t]
\begin{center}
  \caption{KLOE-2 statistical 
sensitivities on several parameters.
}
  \label{tab:modes1}
  \begin{tabular}{|c|c|c|c|c|}
    \hline
\textbf{$f_1$} & \textbf{$f_2$} & \textbf{Parameter} & \textbf{Best published meas.} 
& \textbf{KLOE-2 (50 fb$^{-1}$)} \\
    \hline
$K_S\rightarrow\pi e \nu$      &  
  & 
$A_S$ &
$(1.5\pm 11)\times 10^{-3}$ &
$\pm\,1 \times 10^{-3}$ \\
    \hline
$\pi^+\pi^-$ &  
$\pi l \nu$  & 
$A_L$ &
$(3322\pm 58 \pm 47 )\times 10^{-6}$ &
$\pm\,25 \times 10^{-6}$ \\
    \hline
$\pi^+\pi^-$ &  
$\pi^0\pi^0$   &
$\Re{{\epsilon^{\prime}\over{\epsilon}}}$ &
$(1.65 \pm 0.26) \times 10^{-3}$ (PDG fit)&
$\pm\,0.2 \times 10^{-3} $ \\
    \hline
$\pi^+\pi^-$ &  
$\pi^0\pi^0$   &
$\Im{{\epsilon^{\prime}\over{\epsilon}}}$ &
$(-1.2 \pm 2.3) \times 10^{-3}$ (PDG fit)&
$\pm\,3 \times 10^{-3} $ \\
    \hline
%
$\pi^+ l^- \bar{\nu}$ & 
$\pi^- l^+ \nu$  &
$(\Re\delta+\Re x_-)$ & 
$\Re\delta=(0.25 \pm 0.23) \times 10^{-3}$ (PDG)&
$\pm\,0.2 \times 10^{-3} $ \\
& &  & $\Re x_-=(-4.2 \pm 1.7) \times 10^{-3}$ (PDG)& \\
    \hline
$\pi^+ l^- \bar{\nu}$ & $\pi^- l^+ \nu$  &
$(\Im\delta + \Im x_+)$ &
$\Im\delta=(-0.6 \pm 1.9) \times 10^{-5}$ (PDG)&
$\pm\,3 \times 10^{-3} $ \\
& & & $\Im x_+=(0.2 \pm 2.2) \times 10^{-3}$ (PDG)& \\
    \hline
$\pi^+\pi^-$ &  
$\pi^+\pi^-$  & 
$\Delta m$ &
$5.288\pm 0.043 \times 10^9 s^{-1}$ &
$\pm\,0.03 \times 10^9 s^{-1}$ \\
    \hline
$\pi^+\pi^-$ &
$\pi^+\pi^-$  & 
$\zeta_{SL}$ & 
 $(1.8\pm 4.1) \times 10^{-2}$ & 
$\pm\,0.2 \times 10^{-2} $ \\ 
    \hline
$\pi^+\pi^-$ &
$\pi^+\pi^-$  & 
$\zeta_{0\bar{0}}$ & 
$(1.0\pm 2.1)\times 10^{-6}$ & 
$\pm\,0.1 \times 10^{-6} $\\
    \hline
$\pi^+\pi^-$ &
$\pi^+\pi^-$ & 
$\alpha$ & 
$(-0.5\pm2.8)\times 10^{-17}$ GeV & 
$\pm 2\,\times 10^{-17}$ GeV\\
    \hline
$\pi^+\pi^-$ &
$\pi^+\pi^-$  & 
$\beta$ & $(+2.5\pm2.3)\times 10^{-19}$ GeV & 
$\pm\,0.1 \times 10^{-19}$ GeV\\ 
    \hline
$\pi^+\pi^-$ &
$\pi^+\pi^-$  & 
$\gamma$ & $(+1.1\pm2.5)\times 10^{-21}$ GeV & 
$\pm\,0.2 \times 10^{-21}$ GeV\\ 
& & & & (compl. pos. hyp.) \\
& & & & $\pm\,0.1 \times 10^{-21}$ GeV \\
    \hline
$\pi^+\pi^-$ &
$\pi^+\pi^-$  & 
$\Re \omega$ & $(1.1 ^{+8.7} _{-5.3} \pm 0.9)\times 10^{-4}$ & 
 $\pm\,2 \times 10^{-5} $ \\ 
    \hline
$\pi^+\pi^-$ &
$\pi^+\pi^-$  & 
$\Im\omega$ & $(3.4  ^{+4.8} _{-5.0} \pm 0.6) \times 10^{-4}$  &
 $\pm\,2 \times 10^{-5} $ \\
    \hline
$K_{S,L}\rightarrow\pi e \nu$      &  
  & 
$\Delta a_0$ &
 (prelim.: $(0.4\pm 1.8)\times 10^{-17}$ GeV) &
$\pm\,2 \times 10^{-18}$ GeV \\
    \hline
$\pi^+\pi^-$ &
$\pi^+\pi^-$ &
$\Delta a_Z$ &
 (prelim.: $(2.4\pm 9.7)\times 10^{-18}$ GeV) &
$\pm\,7 \times 10^{-19}$ GeV \\
    \hline
$\pi^+\pi^-$ &
$\pi^+\pi^-$ &
$\Delta a_X$, $\Delta a_Y$ &
(prelim.: $<9.2\times 10^{-22}$ GeV) & 
$\pm\,4 \times 10^{-19}$ GeV \\
    \hline
\end{tabular}
\end{center}
\end{table*}

\section{Conclusions}
The neutral kaon system constitutes an excellent laboratory
for the study of the $CPT$ symmetry and the basic principles of quantum 
mechanics.
Several parameters related to possible $CPT$ 
violations, including decoherence and Lorentz symmetry breaking effects, 
have been measured,
in some cases with a precision
reaching the interesting Planck scale region.
Simple quantum coherence tests have been also performed.
All results are consistent with no violation of the $CPT$ symmetry
 and/or 
quantum mechanics.
\par
A $\phi$-factory represents a unique opportunity to push forward
these studies.
It is also 
an ideal place to investigate the entanglement 
and correlation properties of the produced $\kn\knb$ pairs.
A proposal for continuing the KLOE physics program (KLOE-2) 
at an improved DA$\Phi$NE machine, 
able to deliver an integrated luminosity up to $20\div 50\hbox{ fb}^{-1}$, 
has been recently presented.
Improvements by about one order of magnitude in almost all present limits
are expected.

\begin{acknowledgments}
I would like to thank the organizing committee, and in particular
Elisabetta Barberio and Antonio Limosani
for the organization
of this very interesting and successful conference, 
and the pleasant stay in Melbourne.
\end{acknowledgments}

\bigskip 

\end{document}